# COVID-19 Public Opinion and Emotion Monitoring System Based on Time Series Thermal New Word Mining

**Yixian Zhang[1], Jieren Chen[2,*] , Boyi Liu[3,*] , Yifan Yang [错误!未定义书签。] , Haocheng Li[2], Xinyi Zheng[2], Xi Chen[2], Tenglong Ren[4] and Naixue Xiong[5]**

**Abstract:** With the spread and development of new epidemics, it is of great reference value to identify the changing trends of epidemics in public emotions. We designed and implemented the COVID-19 public opinion monitoring system based on time series thermal new word mining. A new word structure discovery scheme based on the timing explosion of network topics and a Chinese sentiment analysis method for the COVID-19 public opinion environment are proposed. Establish a "Scrapy-Redis-Bloomfilter" distributed crawler framework to collect data. The system can judge the positive and negative emotions of the reviewer based on the comments, and can also reflect the depth of the seven emotions such as Hopeful, Happy, and Depressed. Finally, we improved the sentiment discriminant model of this system and compared the sentiment discriminant error of COVID-19 related comments with the Jiagu deep learning model. The results show that our model has better generalization ability and smaller discriminant error. We designed a large data visualization screen, which can clearly show the trend of public emotions, the proportion of various emotion categories, keywords, hot topics, etc., and fully and intuitively reflect the development of public opinion.

## 1 Introduction

Since December 2019, COVID-19 has been a hot topic for people to discuss on the Internet. In this era of highly developed networks, the open nature of the network and the democratic atmosphere have promoted the development of online public opinion. The situation of epidemic public opinion can affect major industries such as commerce, medical care, transportation, education, and even contribute to scientific research. Analyzing the epidemic situation and public opinion scientifically and systematically to trace the development of public opinion and analyze the changes in public sentiment. It can serve individuals, governments, and even countries. Providing a reliable reference when weighing the "pros and cons" can also predict the key events that can detonate public opinion in the future, so analyzing and visualizing epidemic public opinion data is an indispensable task. Traditional analysis systems include public opinion communication analysis, public opinion feedback analysis, and public opinion subject analysis. The traditional public opinion analysis system has two directions, content analysis, and empirical analysis.

---

[1] School of Information and Communication Engineering, Hainan University, Haikou, 570100, China.

[2] School of Computer Science and Cyberspace Security, Hainan University, Haikou, 570100, China.

[3] University of Chinese Academy of Sciences, Shenzhen, 518000, China.

[4] National University of Defense Technology, Changsha, 410000, China.

[5] Department of Mathematics and Computer Science, Northeastern State University, Tahlequah, OK, USA. Email: xiong31@nsuok.edu.

* Corresponding Author: Jieren Chen. Email: cjr22@163.com; Boyi Liu. Email: by.liu@ieee.org.

Scientifically and systematically analyzing the dynamic development of public opinion from the perspective of the epidemic has the following significance: (1) It can assist the government to implement precise policies and alleviate social panic. During times of public crises, governments must act swiftly to communicate crisis information effectively and efficiently to members of the public [Chen, Min, Zhang et al. (2020)]. The government can implement soothing measures based on changes in public sentiment. Only by keeping abreast of public sentiment dynamics can be the level of public opinion guidance be strengthened; (2) Formulate appropriate isolation measures for the changes in public sentiment, reduce social problems, and reduce various conflicts caused by public unrest. For example, family conflicts, doctor-patient relationship conflicts, and business conflicts caused by emotional unrest; (3) Formulate corresponding economic policies stimulus emotional changes to promote economic recovery. It can also adjust the market according to the changes in people's emotions and demand relationships to achieve the purpose of promoting consumption; (4) It can provide a basis for universities to formulate school opening policies and adjust the school opening time; (5) These findings can help the government and health departments better communicate with the public on health and translate public health needs into practice to create targeted measures to prevent and control the spread of COVID-19 [Zhan, Cheng, Yu et al**.** (2020)].

Therefore, we designed and implemented a COVID-19 public opinion and emotion monitoring system. The system is divided into three modules: text mining, text cleaning and sentiment analysis. In terms of text mining, we use the “Scrapy-Redis-Bloomfilter” distributed crawler framework to crawl corpus, and take Weibo comments as experimental objects. In terms of text cleaning, the system will automatically assemble the corpus in the database and complete the corresponding cleaning work. In terms of sentiment analysis, we try to construct a Chinese word segmentation dictionary using a new idea. In order to make up for the lack of effectiveness of the emotion dictionary in recognizing “infinitive adjective sentences”, we designed a set of preparatory schemes for emotion mapping, and considered the influence of adverbials in the sentence on emotion expression. Finally, we converted the research results into interactive charts and hosted them on the big data visualization platform.

## 2 Related Work

### *2.1 Data crawling*

Obtaining effective data from the Internet is an important part of data analysis. Pais et al. [Pais, Cordeiro, Martins et al. (2019)] develop a specific social network crawler based on API interaction. The APIs provided by the social network can easily and conveniently obtain structured data. So Naskar et al. [Naskar, Singh, Kumar et al. (2020)] use the Search-API provided by Twitter to analyze the emotional dynamics of users' comments on an event or topic on Twitter. Unfortunately, concerning some social networks, such as Facebook, Twitter, Weibo, etc., limited access token dates, API data content, and access times can significantly hamper data collection. Crawler techniques using web crawling have no such limitations. Li et al. [Li, Xu, and Pan (2018)]'s approach is based on Python's selenium to launch the browser, simulate login, analyze the page, and finally get the data from Weibo. Selenium is a great way to deal with the complex login mechanism involved in crawling Weibo data. However, as a suite of tools for automating web browsers, it is not ideal for crawling large amounts of data. Faced with the low data capture rate of a single crawler, Yin et al. [Yin, He and Liu (2018)] propose an improved distributed scheme. With the good asynchronous capability of Scrapy, this scheme introduces the distributed components of the Scrapy-Redis and Redis database into the Scrapy framework and establishes a more efficient distributed crawler system. The data acquisition part of the COVID-19 epidemic opinion analysis system also uses the Scrapy-Redis distributed framework as the main workflow, and the Bloom Filter, a deduplication algorithm based on binary vectors and hash functions, filters and processes the URLs before and after fetching.

### *2.2 Sentiment analysis*

This paper mainly uses text sentiment analysis for data processing, which is a hot research direction of NLP. Its purpose is to mine emotional tendency by analyzing some subjective texts [Li, Jin, and Quan (2020)]. There are two research methods commonly used in text emotion analysis: machine-based learning and dictionary-based [Koupaei, Song, Kristen et al. (2020)]. It is to extract some positive and negative emotion texts from texts as training sets, and classify all texts in positive and negative directions according to the obtained emotion classifiers. The final classification can be a probability value or a 0 or 1 category given for texts. Although this method has been applied in many fields, there are still some shortcomings: (1) It needs manual annotation of training model data, which is time-consuming and laborious. (2) Due to the flexible structure and usage of Chinese sentences, there are many interference factors in feature selection, such as the specialized thesaurus, syntactic structure, information entropy [Shaukat, Zulfiqar, Xiao et al. (2020)]. (3) When the scale of text data is large, there is no guarantee of high accuracy. In contrast, if we can get a large and high-quality emotional dictionary, then we can get more satisfactory results by combining the corresponding semantic rules [Kumar and Babu (2020)].

The key point of emotion classification based on the dictionary is the construction of the dictionary. At present, domestic and foreign scholars have made some achievements. Nie et al. [Nie, Tian and Wang et al. (2020)] built a new hotel selection model based on online text review on tripadvisor.com, put forward a semantic mapping function and a method of building a dictionary, and finally verified the validity of the model through case analysis, robustness analysis, and comparative analysis. Chen et al. [Chen, Li, Wang et al. (2020)] proposed a learning framework based on multi-source information fusion (MSIF), extracted four kinds of emotional information: lexical emotional information, emotional word co-occurrence information, emotional word polarity information and emotional word polarity relationship information, and solved the construction problem of domain emotional dictionary. Ahmed et al. [Ahmed, Chen, and Li (2020)] proposed an attention-based LSTM model to deal with the aspect level emotion analysis task of emotion score retrieved from the proposed dictionary. In a specific sentence, the non-emotional part of the aspect related information is weighted, and the problem of polarity detection in the dictionary is effectively improved. To sum up, although the construction of domain dictionaries is relatively mature, there are not many involved in the construction of basic dictionaries, or the scale of the selected basic dictionaries is too small. Due to the different application scenarios, we need to select the appropriate basic emotion dictionary and build a reasonable field emotion dictionary. This research integrates resources such as Tsinghua commendatory and derogatory dictionary, Weibo public opinion dictionary, Dalian University of technology emotion ontology vocabulary, uses SO-PMI mutual information algorithm to build an emotion dictionary related to COVID-19, which applies to the network language environment, and subdivides the emotional particles to analyze seven emotional results. It realizes the dynamic multi-angle visualization of the changes of public emotion under the epidemic situation, which can be helpful for the grasp of public psychology and the prevention and control of the epidemic situation.

## 3 Methodology

### *3.1 Text mining*

#### *3.1.1 Crawler frame selection*

The "Scrapy-Redis-Bloomfilter" distributed crawler framework is the main workflow of capturing comments [Deng, Liu, and Dong (2018)]. When the number of crawling levels exceeds 100 million, Redis will occupy a lot of space. At the same time, Redis will also occupy memory due to the storage of crawling queues, which makes it difficult to achieve simultaneous crawling of Scrapy projects. Here we use a more memory-saving deduplication algorithm Bloom Filter, which has huge advantages in space and time. Bloom Filter first uses a bit array to represent a set to be detected, and can quickly determine whether an element exists in this set through a probabilistic algorithm, to achieve the deduplication effect.

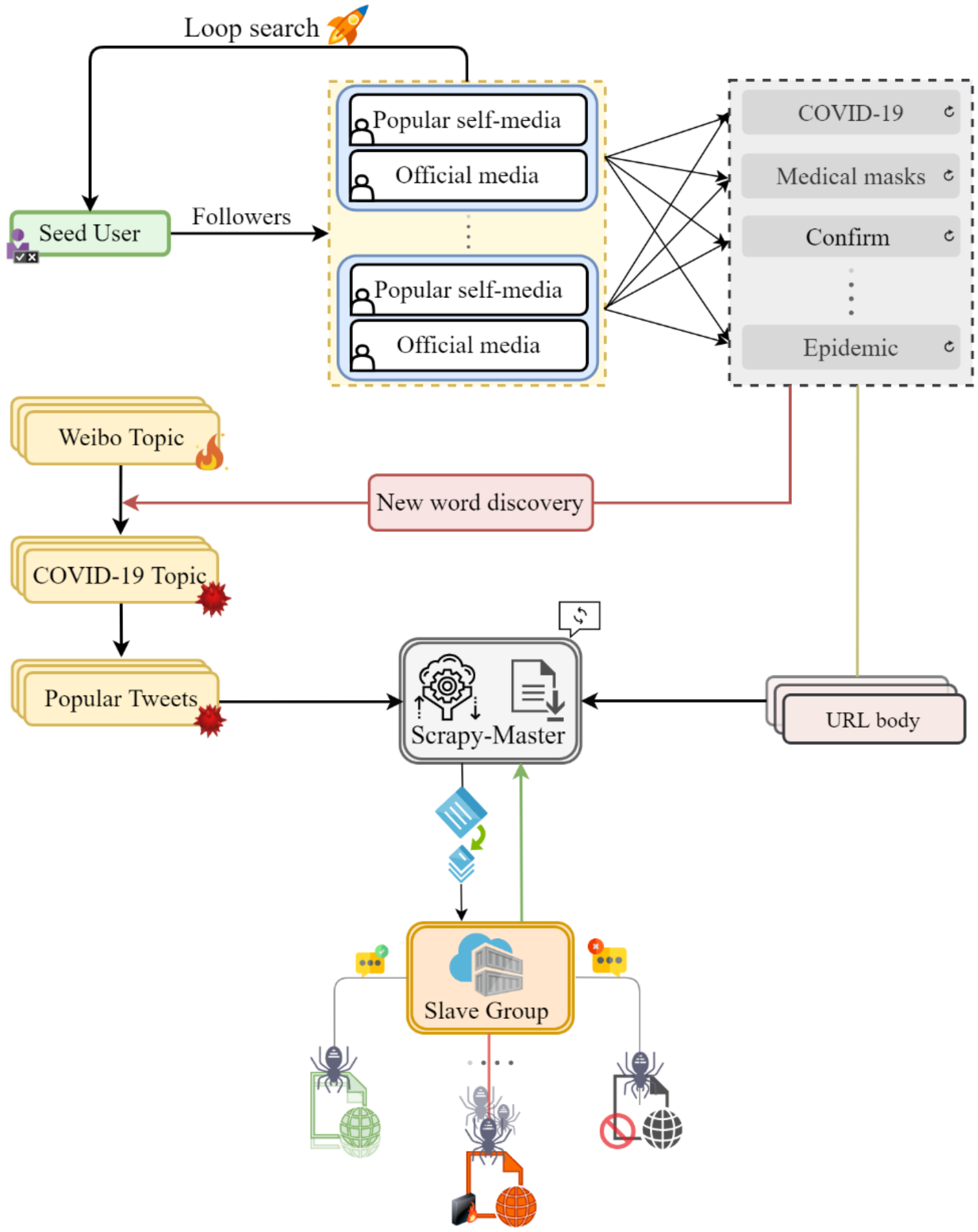

**Figure 1:** The "Scrapy" distributed crawler framework

*3.1.2 Anti-anti-reptile technology*

To fully respond to the more flexible anti-acquisition strategy of the hot search of Weibo, we have prepared multiple sets of solutions to counter the detection of anti-reptiles on Weibo to ensure cloud acquisition tasks. The specific plan is as follows:

(1) Rewrite the proxy IP pool maintenance script. Monitor the websites of several major proxy IP providers in China, use the "dual process + multi-thread multi-coroutine" maintenance mode, and operate the "proxy IP pool verification" and "proxy IP pool rotation" asynchronously to ensure that the available IP survival rate is 90%. The above guarantees that more than 100 HTTPS-type proxy IPs can be used by the main crawler framework in real-time. The use frequency of proxy IP is controlled at 1/5 minutes.

(2) Monitor the homepages of 24 domestic network service providers based on the STAFF

framework, collect the relay server IP; the distributed crawler carries Socks5 to disguise HTTPS traffic for global access.

(3) Prepare 200 micro-blog verification-free accounts, regularly simulate login cracking verification to ensure that the cookie pool is available.

#### *3.1.3 Data analysis*

We have chosen two methods for data analysis to ensure that most useless information is filtered.

Priority call the crawler intelligent analysis library newspaper, compared to readability, the function provided by the newspaper library is more powerful. First of all, the Article class is first imported into the newspaper library, directly passed in the URL, and its download method is called. Then we execute the parse method to intelligently parse the web page. Finally, according to the needs of the project, intelligently filter useless data.

In Scrapy, there are two ways to extract data, one is Xpath selector and the other is CSS selector. In the Scrapy crawler framework, the text () function is often used together with Xpath expressions to extract the data content of the node. Therefore, in this experiment, we use Scrapy's built-in parser Scrapy Selector's XPath tool to parse HTML information.

### ***3.2 Data cleaning***

Data cleaning, also called data cleansing or scrubbing, which means detecting and removing errors or inconsistencies from data to improve the quality of data [Wang, Wang, Yang et al. (2020)]. The main goal of data cleaning is to prepare for the next sentiment analysis, and segmentation of phrases is the key to data preprocessing.

The word segmentation method is based on the word frequency statistics of the existing corpus. For large-scale texts obtained from Weibo, Out of Vocabulary has a greater impact on the accuracy of word segmentation than ambiguous segmentation, especially network new words from Weibo are often not registered in the corpus. Hidden Markov model plays an important role in natural language processing (NLP) for PART-OF-SPEECH (POS) tagging. Nambiar et al. built a Hidden Markov model by using the existing marker sentences in Malayan language [Nambiar, Leons, Jose et al. (2019)]. The Viterbi algorithm based on the HMM model predicts the optimal hidden state sequence satisfying the observation sequence by giving the known observation sequence and model parameters.

(1) According to the different components of the word in the phrase, create a tuple. The tuple containing four status is Status [3] = [B (located at the beginning), E (located at the end), M (located between the phrases), S (separately formed words)]; Set a Weight and Path for each Chinese character (Weight: under a certain state, the probability of occurrence of the word, Path: represents the state of the previous word when the emission probability of a certain word reaches the maximum value).

(2) Use the Viterbi algorithm to calculate the weight of the entire text

Implicit state transition probability matrix:

$$\text{Observed}[i-1]\ \begin{matrix} B \\ E \\ M \\ S \end{matrix} \begin{bmatrix} P_{11} & P_{12} & P_{13} & P_{14} \\ P_{21} & P_{22} & P_{23} & P_{24} \\ P_{31} & P_{32} & P_{33} & P_{34} \\ P_{41} & P_{42} & P_{43} & P_{44} \end{bmatrix} \text{Observed}[i] \tag{1}$$

Observed [i] represents the i-th font of the observed text sequence, Observed [i-1] represents the i-1th font; according to the limited historical assumption of the HMM model, where $P_{ij}$ = P (Observed [i] | Status [j]), indicating the probability of Observes [i] appearing in a certain state.

Calculate the weight [Observed [0]] of the first word in four states:

Weight (Observed [0]) = P (Observed [1], Status [i]) = P (Status [i]) * P (Observed [0] | Status [i]);

By comparison, the maximum weight of each font is obtained to determine its corresponding status, which is the "optimal Path".

After getting all the "optimal paths", traverse the status value set of the output text according to its backtracking, and segment the phrases according to the different states of the text. The overall method process is shown in Figure 2.

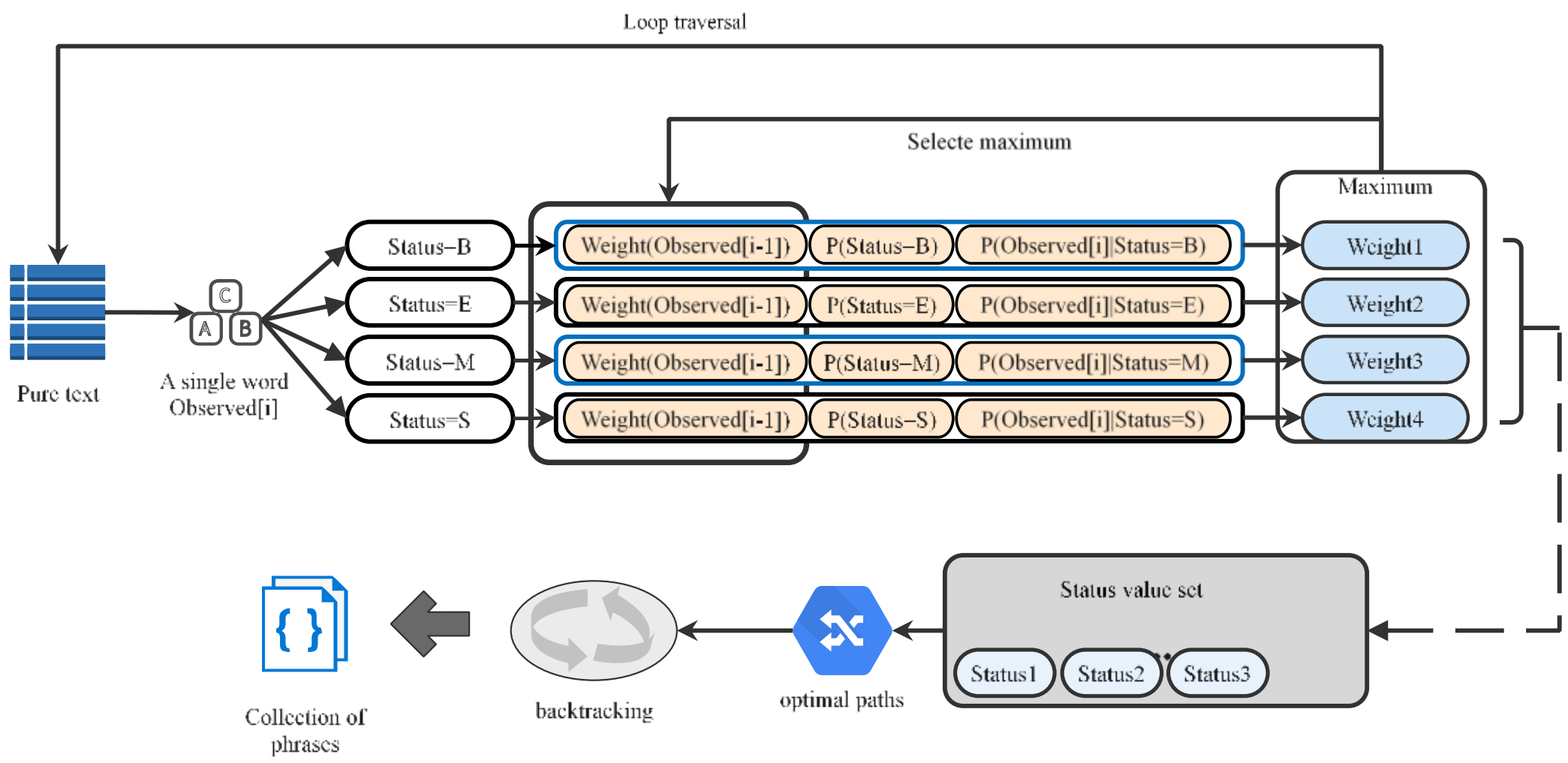

**Figure 2:** The Viterbi algorithm based on the HMM was used to segment the unregistered words

### *3.3 Sentiment analysis*

Compared with other text analysis, the text sentiment analysis part is a relatively difficult one. For the task of text emotion classification, we first extract emotion features from documents, and then use the classifier to classify them [Yi Cai, Kai Yang, Dongping Huang et al. (2019)]. In this study, sentiment dictionaries for the microblog review environment are selected as classification features, and a naive Bayesian method is used to construct a classifier for text sentiment analysis. Finally. we subdivide the emotional particles.

*3.3.1 Establish a microblog comment sentiment dictionary*

The coverage and completeness of the dictionary often determine the final analysis effect. To improve the accuracy of sentiment analysis, we need to expand on the existing Weibo public opinion dictionary [Zhao Jia-Lin, Li Meng-Zhu, Yao Juan et al. (2019)]. The method used in this study is to integrate and select the appropriate open-source sentiment dictionary and use the SO-PMI mutual information algorithm to quickly and easily build a domain sentiment dictionary related to COVID-19 and adapted to the network language environment.

The general sentiment dictionaries we selected are Tsinghua sentiment dictionary, BoSonNLP praise and derogatory dictionary, Dalian University of Technology emotional ontology library. Considering that the network's new word update speed is very fast, we choose to analyze the sampled information of each day based on the time dimension based on these general sentiment dictionaries to discover new words and expand our sentiment dictionaries.

*3.3.2 Text preprocessing*

Text preprocessing includes Chinese word segmentation and removal of stop words. The Chinese word segmentation has been elaborated in the above, so I won't go into detail here. To remove stop words is to traverse all the words in the corpus and delete the stop words. For example: Nanjing has never had such a small number of people on the street, never been so quiet, the dream of working at work without traffic jams, even happened at such a time. After removal: never / no / so / less / never / no / so / quiet / no / actually.

*3.3.3 Build the model*

*3.3.3.1 Construct a naive Bayes classifier to classify words*

Perform a preliminary polarity classification on the word segmentation results to determine whether it is positive or negative. Naive Bayes algorithm is used here.

Naive Bayesian classifier (NBC) assumes that the influence of an attribute value on a given class is independent of the value of other attributes. This assumption is called class conditional independence, and in this sense is called naive. It is an algorithm based on probability, which calculates the posterior probability that the Y variable belongs to a certain category according to certain prior probabilities. The specific steps are: based on the constructed vector matrix, the calculation formula is as follows:

$$C_{NB} = argma\underset{C_j \in C}{X} \left\{ P(C_j) \prod_{i=1}^{n} P(w_i, c_j)^{wt(w_i)} \right\} \quad (2)$$

Among them: $P(C_j)$ is the prior probability of the category; $P(w_i, c_j)$ is the posterior probability of the feature word $w_i$ in the category; $wt(w_i)$ is the weight of the word $w_i$ in the test corpus, when the bool weight is used, $wt(w_i)$=1 .

For the prior probability, this study uses the annotated corpus to estimate in advance, and the calculation formula is as follows:

$$P(c_j) = \frac{doc(c_j)}{\sum_{c_j \in C} doc(c_j)} \quad (3)$$

Where doc($c_j$) is the number of texts belonging to the category $c_j$

For the posterior probability, this study uses the sum of the weight of the word $w_j$ in the text of $c_j$ divided by the sum of the weight of all the words in the text of $c_j$. The specific calculation formula is as follows:

$$P(w_i, c_j) = \frac{weight(w_i, c_j) + \delta}{\sum_{i=1}^{n} weight(w_i, c_j) + \delta |V|} \quad (4)$$

$\left(V = \sum_{c_j \in C} \sum_{i=1}^{n} weight(w_i, c_j), \delta = \frac{1}{|V|}\right)$ Loop through the remaining fonts of the text:

Where $weight(w_i, c_j)$ is the sum of the weights of the word $w_i$ in the text belonging to the category $c_j$.

NBC is also known as the optimal classifier, and it has outstanding performance in text classification applications. When processing large-scale data sets, the classification accuracy is high and the implementation is relatively simple.

*3.3.3.2 Subdivide emotional particles*

The 7-dimensional emotional level is scored through the emotional ontology and custom extended dictionary of the Dalian University of Technology. The higher the score, the greater the proportion of this emotion.

The emotions in the vocabulary ontology are divided into seven types: Hopeful, Happy, Depressed, Angry, Frightened, Disappointed, Shocked, the emotional intensity is: 1, 3, 5, 7, 9, 5 levels, 9 intensity is the largest, 1 intensity is the least. Each word corresponds to information such as polarity under each type of emotion. Turn the above segmentation result into a dictionary.

Classify the results of word segmentation, find emotion words, negative words, and degree adverbs.

First set the initial weight W to 1, start with the first sentiment word, use the weight W * sentiment value of the sentiment word as the score (recorded with score), and then determine whether there are a degree adverb and negative word with the next sentiment word. If there is a negative word, W * -1 if there is a degree adverb, W * degree adverb degree value, W at this time as the weight value of traversing the next emotion word, loop until all emotion words have been traversed, the sum of the scores during each traversal is the sentiment score of this sentence. The predicted value is in the form of a cumulative sum, and the larger the score of the last category the more predicted the category. (In which case there will be negative numbers, you need to find the absolute value before comparing).

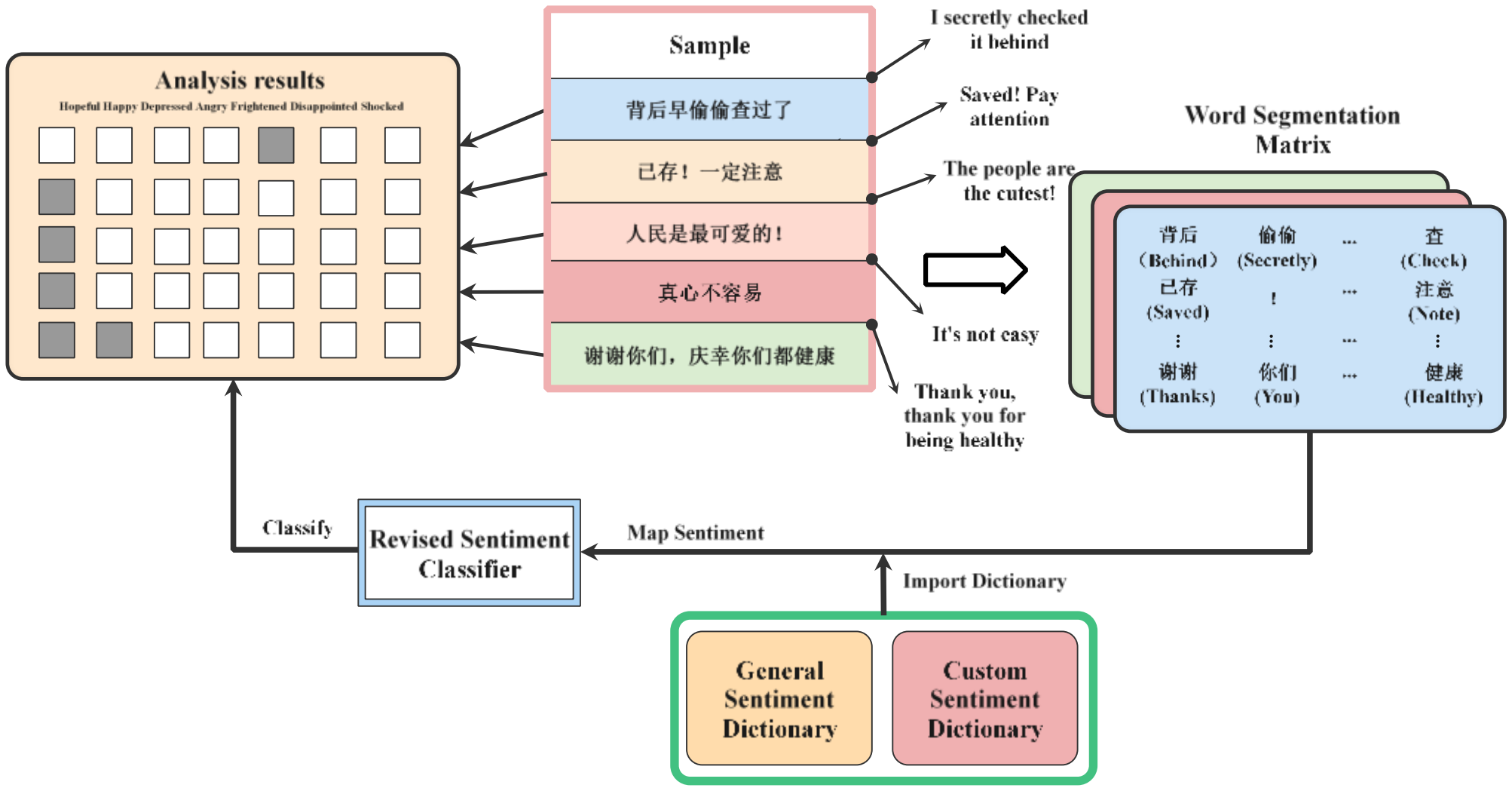

**Figure 3:** Flow chart of sentiment analysis

## 4 Experiment

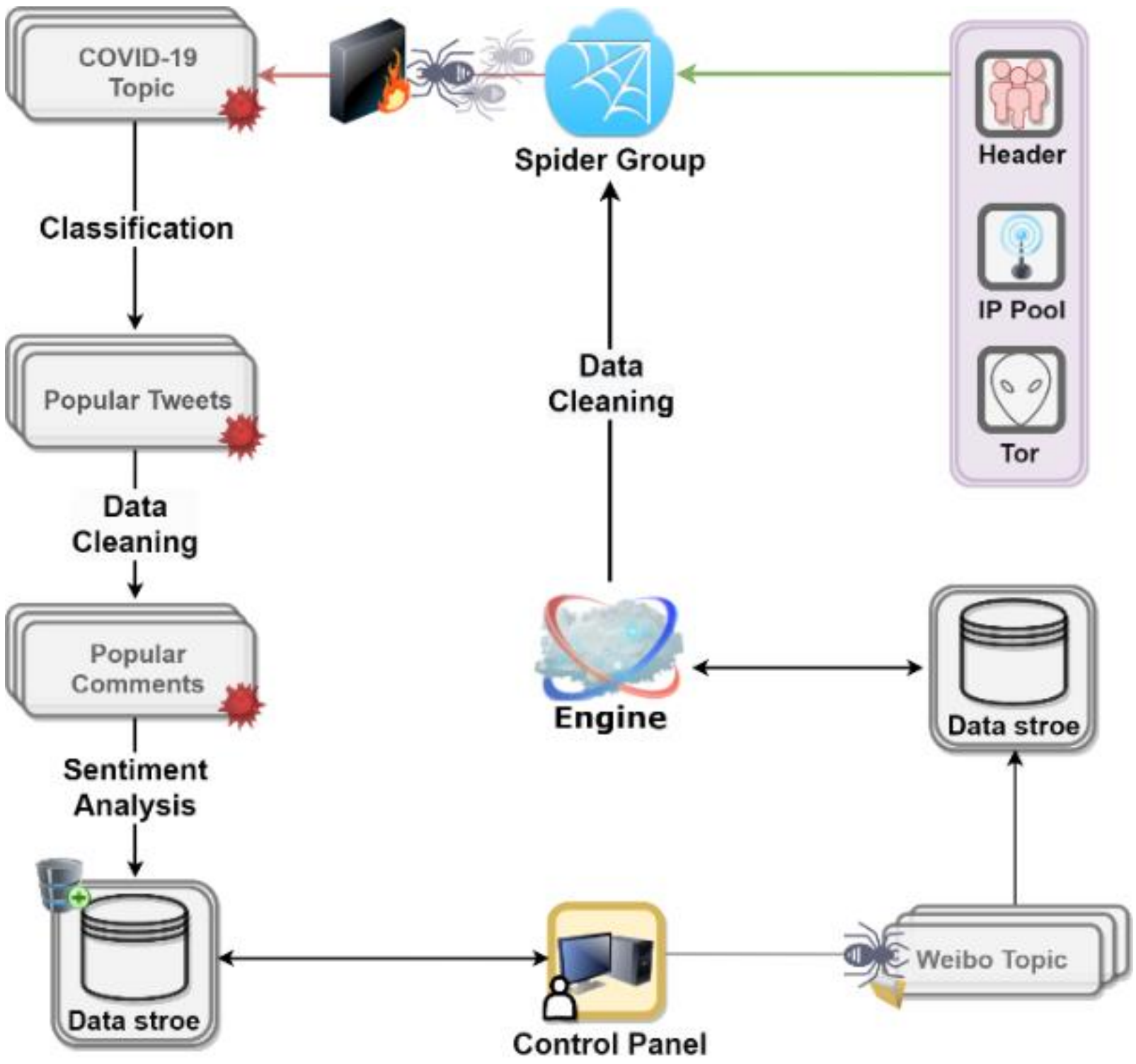

**Figure 4:** The principle framework of COVID-19 public opinion and emotion and emotion monitoring system

### *4.1 Brief description of experimental objects and framework*

Social media represented by Weibo is an important platform for people to express their emotions and feelings. Weibo has the characteristics of the last update, wide information sharing, and rapid spread. Users can obtain and publish information according to their interests and preferences, so choose Weibo Bo is the experimental object.

Through the crawler protocol, we request data on Weibo information about the COVID-19 epidemic, analyze the obtained data, parse the web page, and extract text information. To ensure that the experimental data is more accurate, we take two preprocessing of data cleaning the sentiment analysis of the text will store the obtained experimental data.

### *4.2 Experimental configuration*

In this experiment, we chose a computer with a memory type of DDR4 2400MHz and a hard disk capacity of 128GB SSD + 2TB HDD for the entire experiment. The following table details the experimental PC configuration.

**Table 1:** PC side

| | |
|---|---|
| Memory type | DDR4 2400MHz |
| Hard drive capacity | 128GB SSD+2TB HDD |
| Memory capacity | 4GB |
| Memory type | GDDR5 |
| Memory bit width | 128bit |
| processor | Intel(R) Core(TM) i5-8400 CPU @ 2.80GHz 2.80 GHz |
| RAM | 16.0 GB |
| system version | Windows 10 Home Chinese Edition |
| Graphics chip | NVIDIA GeForce GTX 1050 Ti |

On the server-side, we chose two servers BandwagonHOST and Baidu BCH as experiments. We show the configuration of the experiment server clearly and in detail through a table.

**Table 2:** Server

| Server | BandwagonHOST | Baidu BCH |
|---|---|---|
| Operating system | Centos 7 x86_64/ Python 3.6.6 | Centos 7 x86_64/ NGINX+PHP 5.4 |
| RAM | 2GB | 2GB |
| Disk | 20GB SSD | 10GB SSD |
| SWAP/Data Store | 260MB | MySQL5.5.35/1000M |
| Bandwidth | 2.5Gbps | 2Mbps |

### *4.3 Visualized large screen display*

Compared with traditional charts and data dashboards, today's data visualization is dedicated to presenting business insights hidden behind rapidly changing and complex data in a more vivid and friendly form. Whether in retail, logistics, electricity, water conservancy, environmental protection, or transportation, interactive real-time data visualization large screens help business personnel to discover and diagnose business problems and increasingly become an indispensable part of big data solutions. Therefore, for the part of visually displaying the analyzed data, we chose the Baidu intelligent cloud product Sugar, which uses a large-screen visualization method to analyze and display complex data products. The purpose is to let more people see the charm of data visualization. Interface to build professional-level visualization pages.

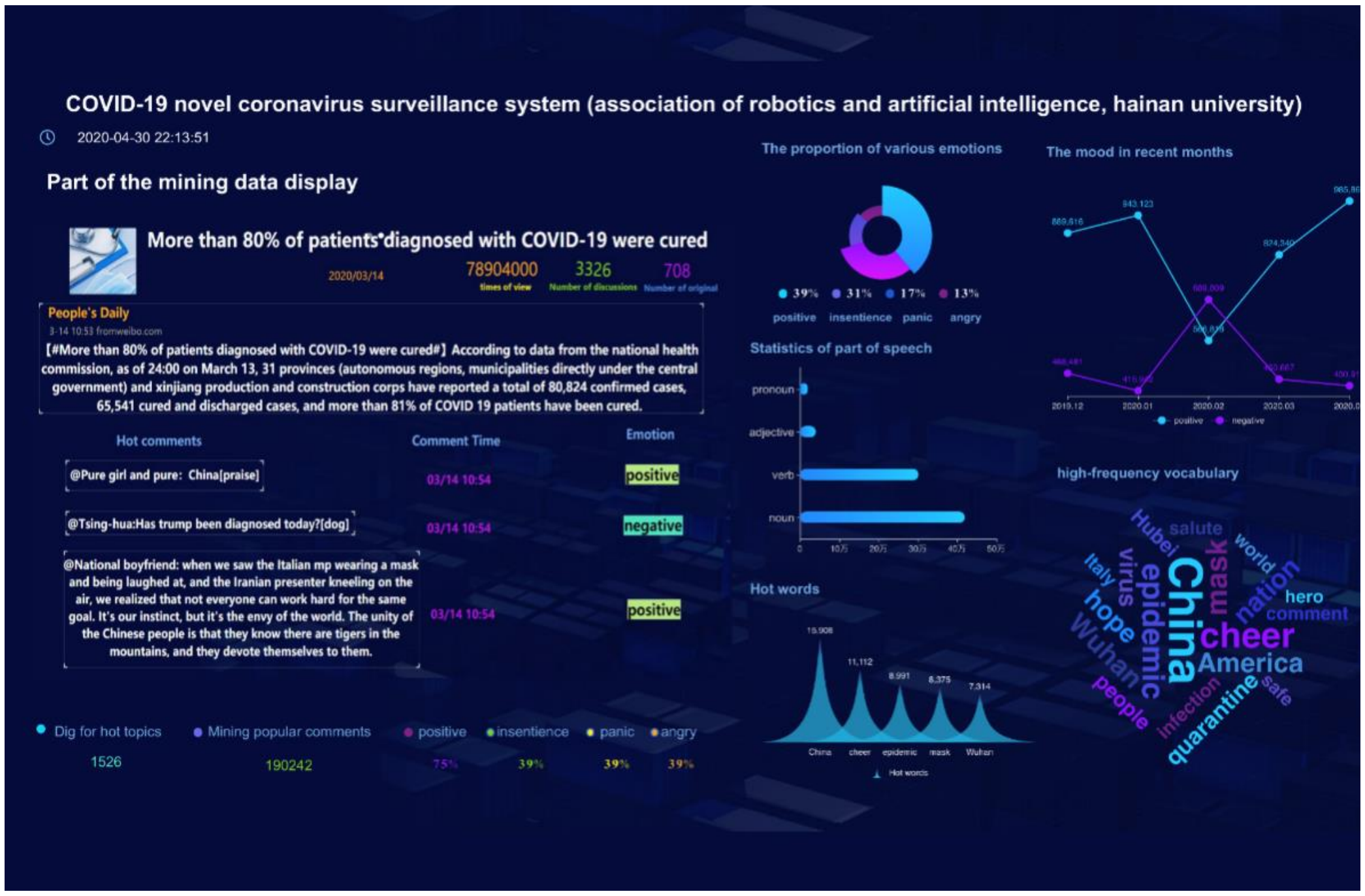

**Figure 5:** COVID-19 epidemic situation public opinion monitoring system visualization big screen

Data display: After removing invalid data through data cleaning, we tapped more than 1,526 hot topics and 190,242 hot comments.

### *4.4 Experimental results display and sentiment analysis*

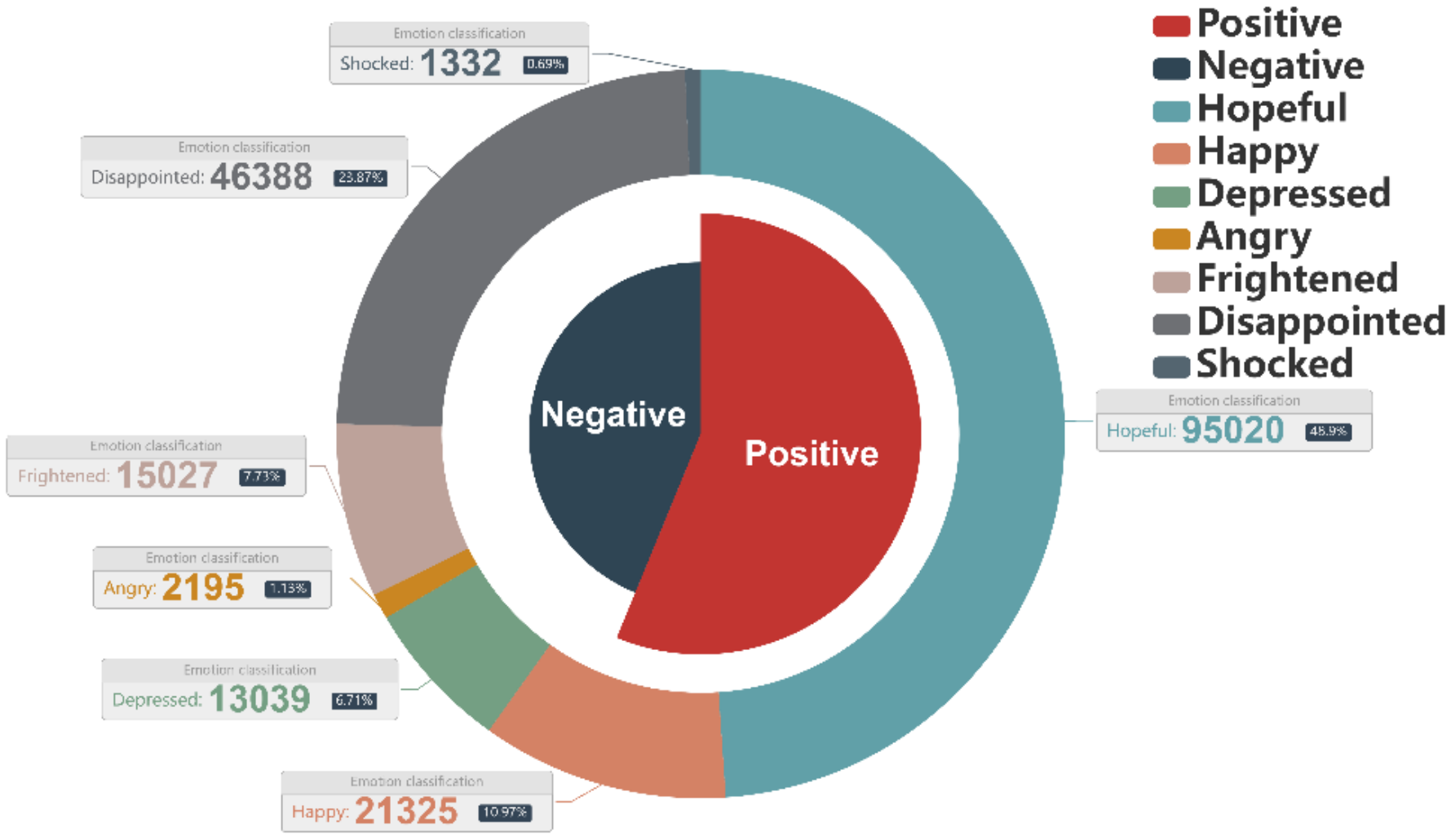

**Figure 6:** Classification of user emotions under the COVID-19 epidemic

There are two kinds of emotions: positive and negative, but positive and negative alone are not enough to express people's deeper emotions. This experiment further refines the emotions of Weibo users and subdivides richness based on positive and negative. Come up with seven detailed emotions: hopeful, happy, depressed, disappointed, angry, frightened and shocked, and through the data analysis of the number of Weibo users and the proportion of each emotion, we can better understand the deep emotions of blog users, and can more accurately determine the user's emotional attitude towards the COVID-19.

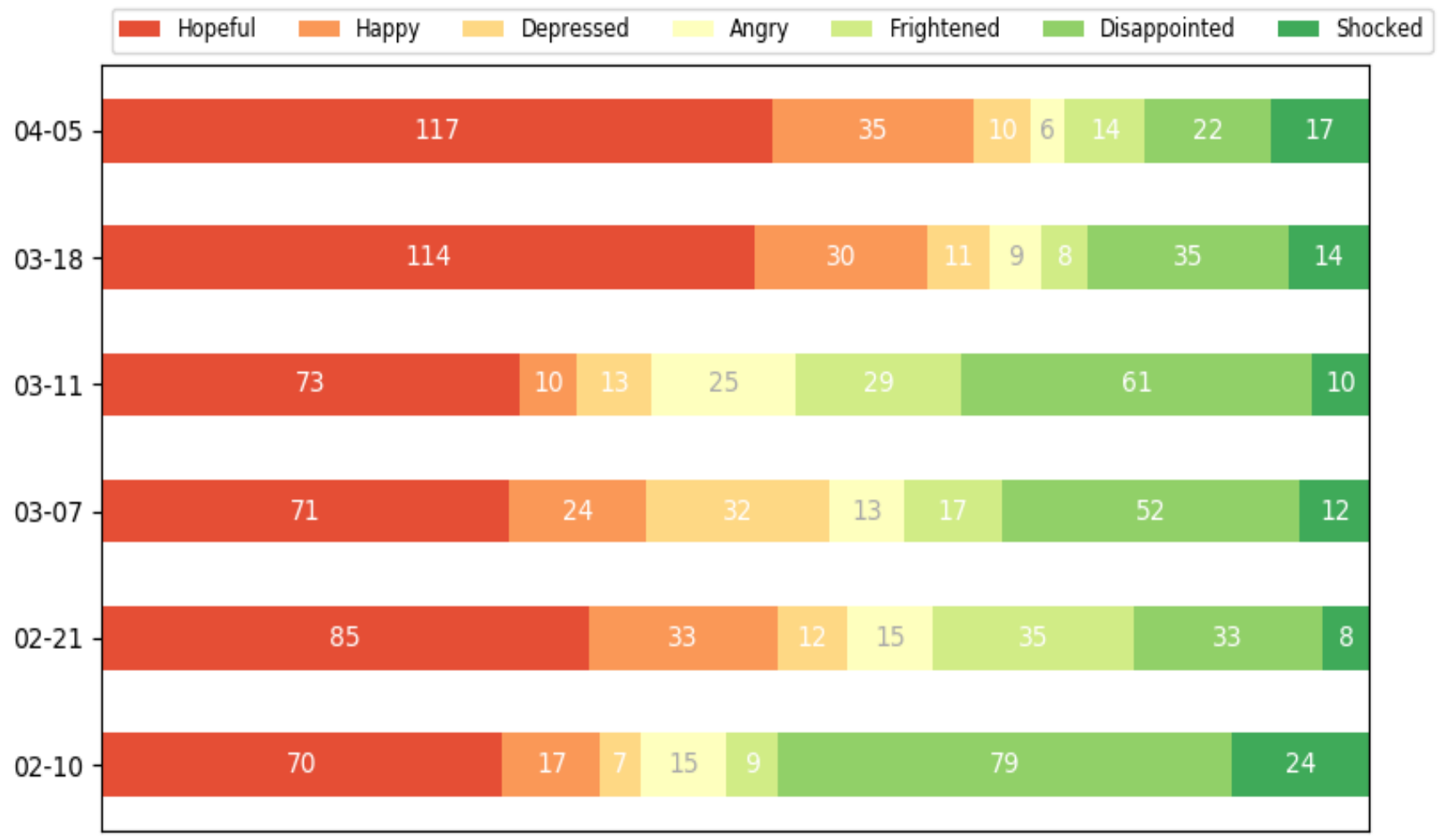

**Figure 7:** Changes in the percentage of user emotion classification within a specified period

Figure 7 shows seven deeper changes in positive and negative emotional cycles. We have analyzed the changes in the emotional cycle from the severe period of the COVID-19 from February to early April 2020, and we can clearly see the changes in seven emotions in each period from the figure. You can grasp and judge public opinion more scientifically and effectively. As shown in Figure 7, we can observe that since mid-February, people have become more and more positive in the face of the COVID-19, and their negative emotions have declined.

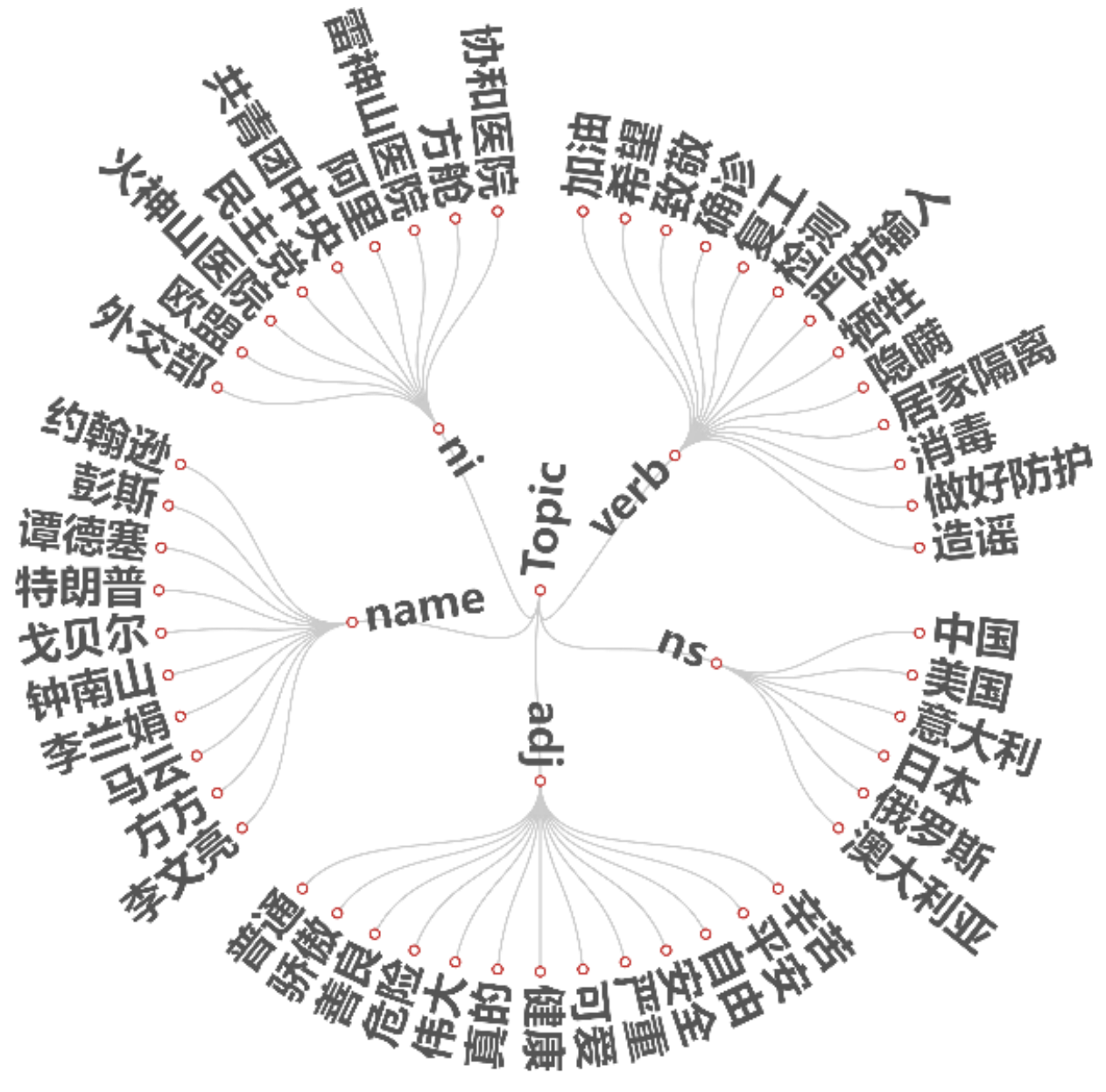

**Figure 8:** Keyword representation by part of speech

Different parts of speech often have different effects. We choose the verbs, adjectives, related names, country names, and institution names of users regarding COVID-19. Adjectives can best express the emotions of Weibo users, such as safety and health; country names can reflect the user's concern about the epidemic situation in various countries of the world; and the name of the person is the focus and hot person in the COVID-19 epidemic situation, which can often drive the direction of public opinion. For example, Professor Zhong Nanshan's speech has a great influence on public opinion; the emergence of hot words from institutions also reflects the public's concern about the epidemic. For example, the construction process of Vulcan Mountain Hospital affects people's hearts.

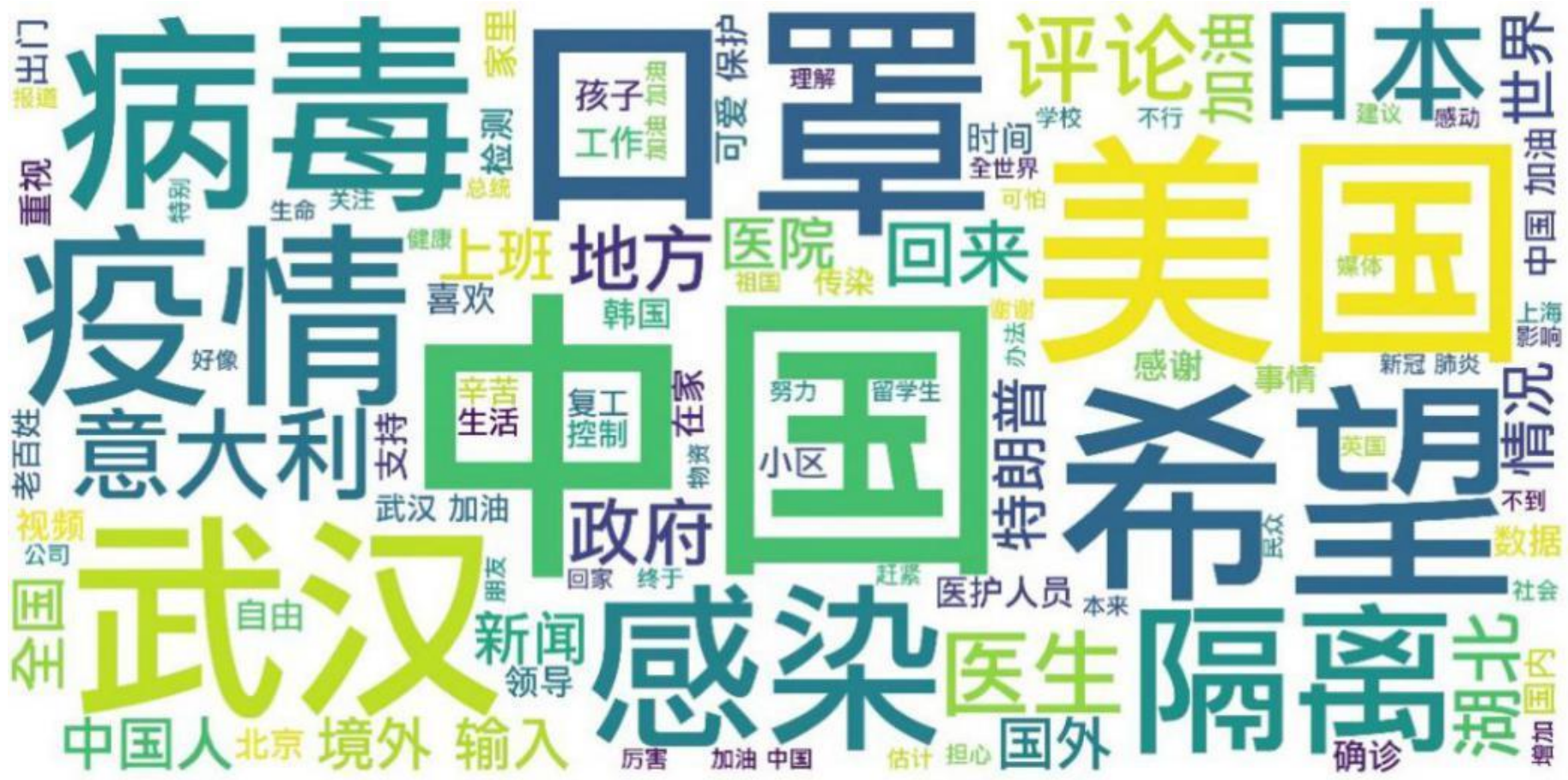

**Figure 9:** Chinese word cloud

The Chinese word cloud generated after the analysis of the text of this experiment. The content of the word cloud includes China, epidemic situation, refueling, virus, isolation, etc.; the word cloud is a visual representation of the text data, which is composed of words and a cloud-like color graphic used to display a large number of texts. data. At the same time, the importance of each word is displayed in font size or color, which allows the viewer to perceive the weight of certain keywords most quickly. Keywords often indicate the trend of public opinion.

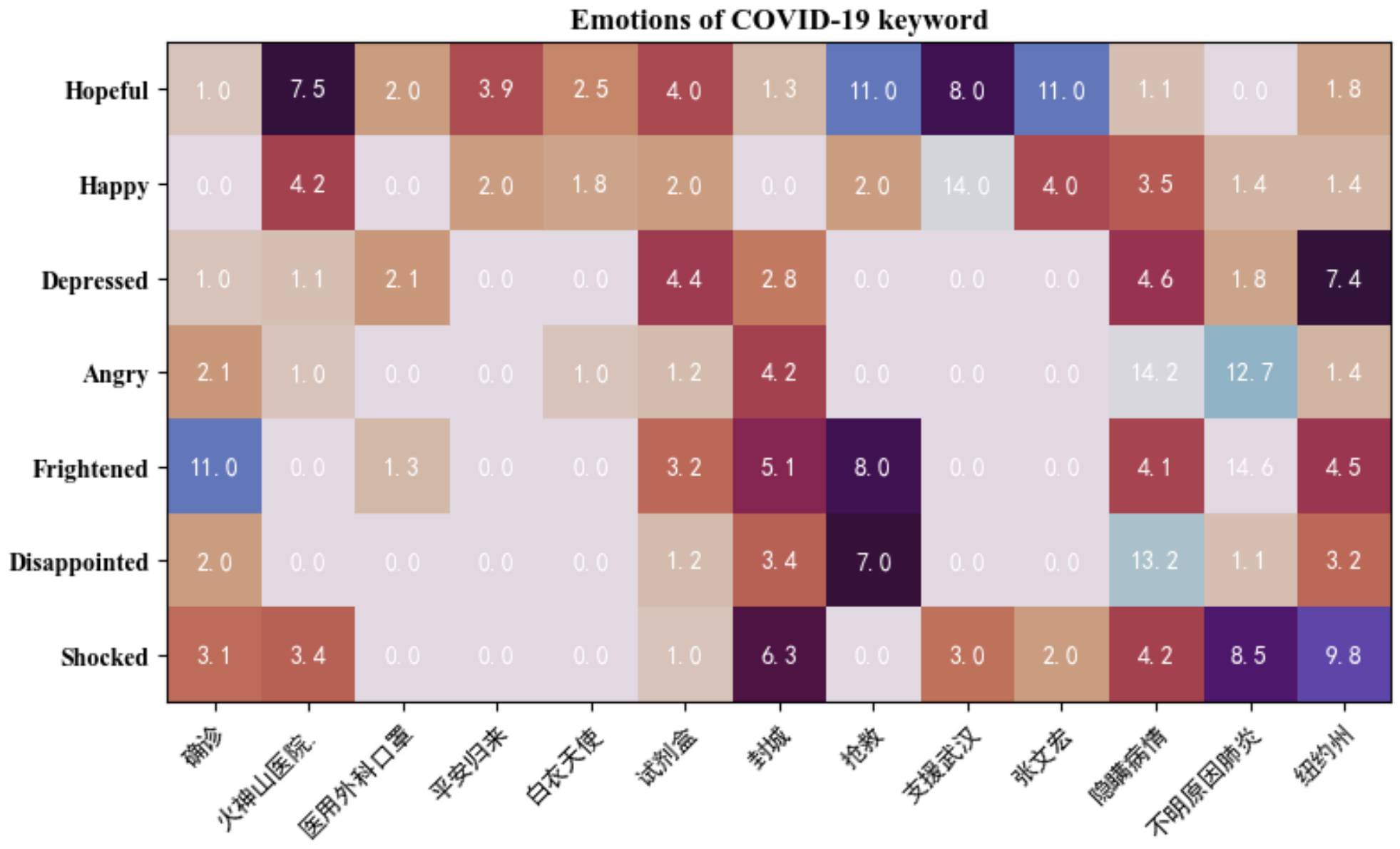

**Figure 10:** Emotional score chart of designated hot words

We randomly selected 13 hot words which Weibo users to search and discuss such as the diagnosis, closed the city, concealed the condition, and return safely and so on, because these hot words can best represent the public's attitudes and sentiments about the COVID-19, and can best reflect public opinion. Because each hot word contains a variety of complex and comprehensive emotions, we use these emotional words to score the seven-dimensional emotional level through the Dalian University of Technology's emotional ontology and a custom extended dictionary. The higher the score, the greater the proportion of this emotion. The significance of this graph is that it can make predictive instructions, such as adding hot words such as workers returning to work, students returning to school, etc., we can understand the user's emotional attitude to this event, so that the relevant departments can make better decisions.

### *4.5 Comparative experiment*

Jiagu deep learning natural language processing tools can perform natural language processing and emotional analysis to a certain extent, but Jiagu natural language processing often has certain errors with the actual language emotion expression. Our modified model can better perform natural language. Sentiment analysis reduces errors and improves the accuracy of sentiment analysis. In this experiment, we randomly sampled 4000 comments of Weibo users after data cleaning, and conducted experiments with Jiagu and our revised model, and compared the experimental results.

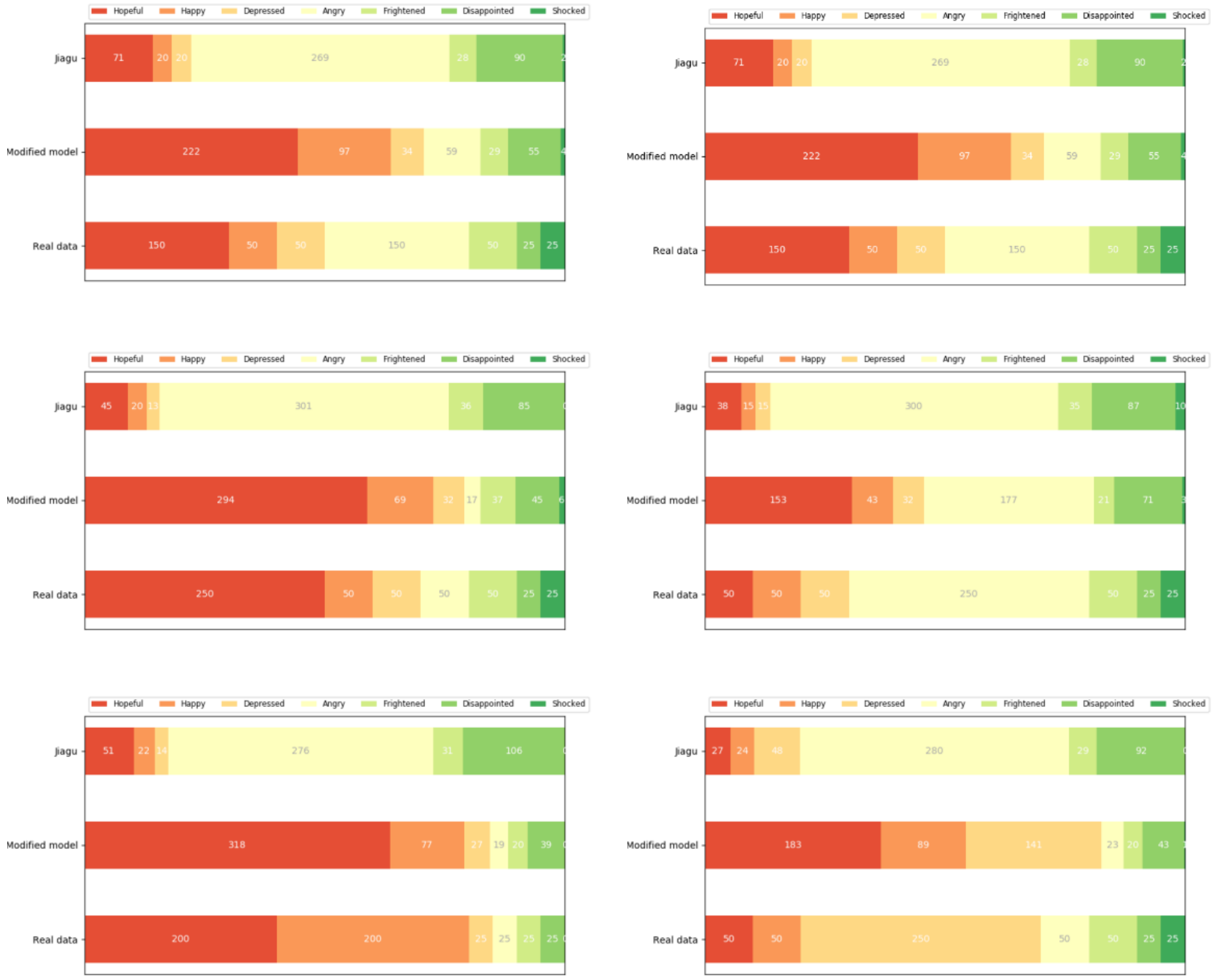

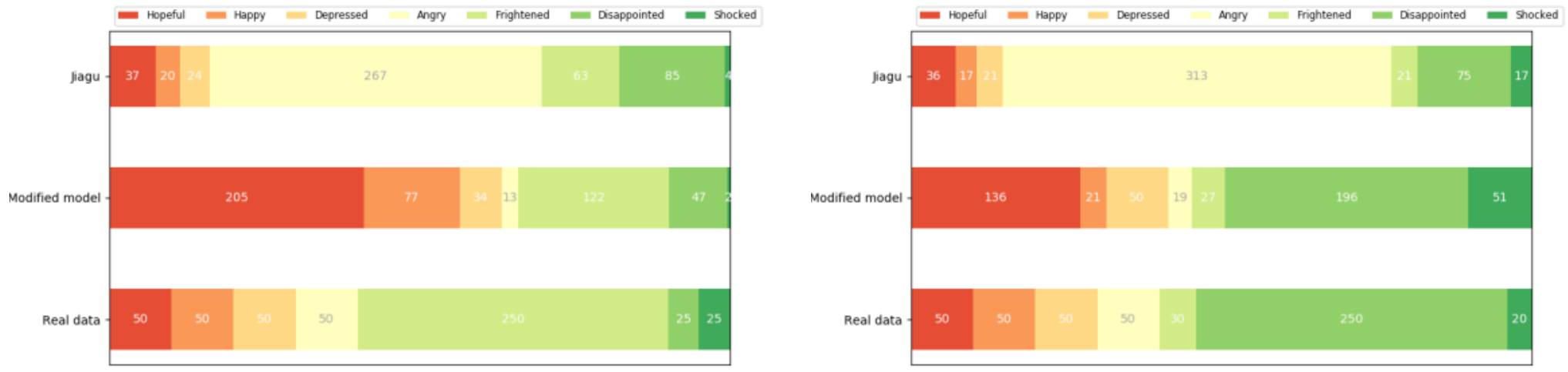

**Figure 11:** Comparison of Jiagu and Modified model with Real data under random sampling

As shown in Figure 11, divide the comments of chosen randomly divided eight samples, using the modified model and Jiagu to deal with each sample and sentiment analysis respectively, with the real data as a reference of artificial judgment, draw eight-bar chart (1, 2, 3 lines represent the Jiagu, correction model and real data). The results show that, within the allowable range of error, the proportion of emotion partition obtained by using the modified model is closer to the real data and the error is smaller than that obtained by using Jiagu.

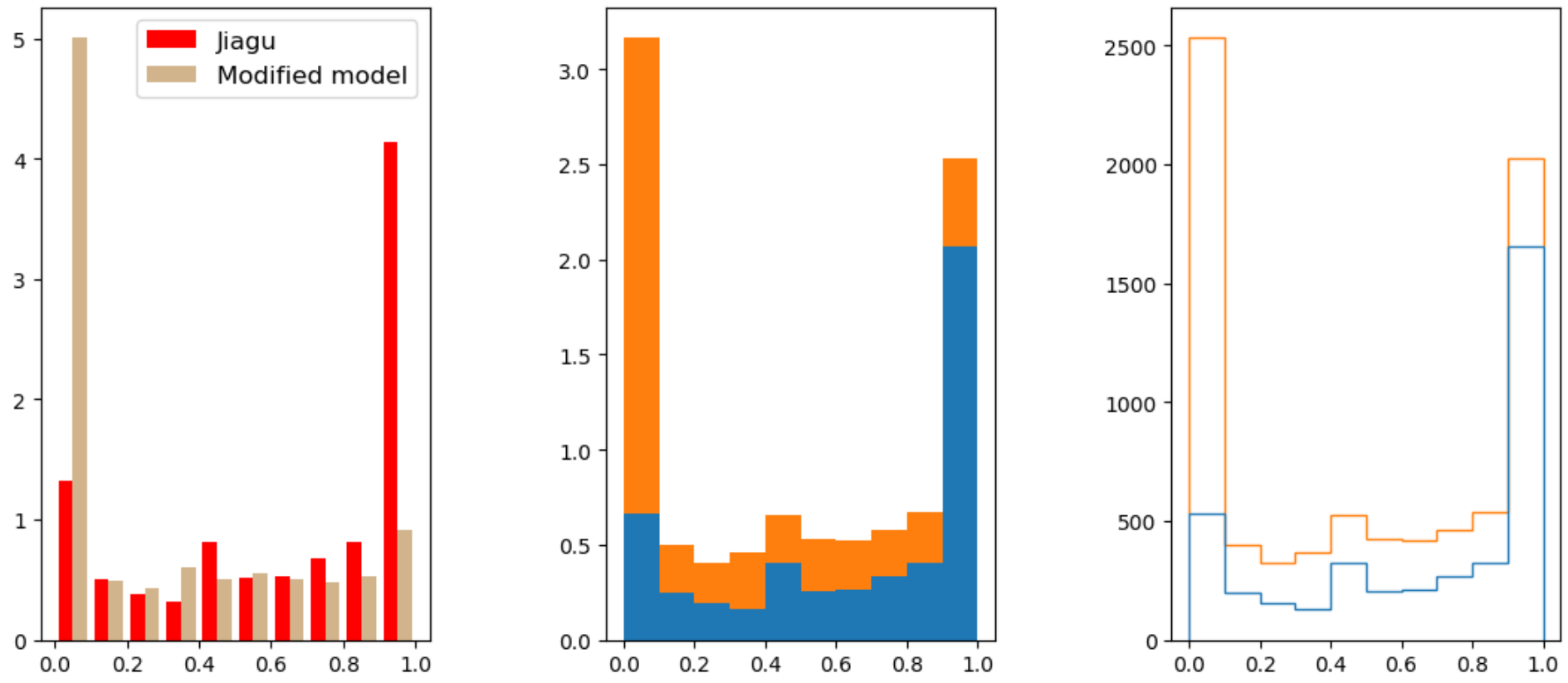

**Figure 12:** Error analysis chart of comparative experiment between Jiagu and Modified model

We use 0.2 as the boundary to divide the error gradient and compare the 1991 positive comments of the modified model with the 2000 positive comments processed by Jiagu. Observe the Figure 12 to judge the error of Jiagu and the modified model. We can see clearly from these three bar charts that the error of the modified model is less than that of Jiagu. It should be noted that since some sentences cannot be commented on for emotional analysis, there is a 0.5% error in this experiment that cannot be judged in the experiment.

## 5 Conclusion and our future work

In this work, we designed the COVID-19 public opinion and emotion monitoring system based on time series thermal new word mining. Our system has the following two innovative attempts. First of all, given the characteristics of COVID-19 public opinion communication, we tried to use an improved dictionary construction method based on the SO-PMI algorithm to help us adapt to the COVID-19 network public opinion environment under the conditions of limited transferable models and insufficient seed corpus. Build a Chinese word segmentation dictionary. Secondly, we have comprehensively absorbed the experience methods of many excellent Chinese sentiment analysis models and designed a set of discriminating processes on this system. When the sentiment dictionary cannot directly and effectively classify the deep emotions of the text, we can classify and discriminate according to the benchmark positive and negative emotion Confidence maps emotions to deep emotions to achieve indirect classification. At the same time, we have integrated some excellent online lexicons and emotional lexicons to further expand the lexical corpus and improve the accuracy of system discrimination. We compare the improved model with the Jiagu deep learning model. The results show that the improved model has a stronger level of generalization and a lower discrimination error. Finally, we designed a large data visualization screen to share the team's research results. The data displayed includes but is not limited to fluctuations in the emotional timing of netizens on related topics, the overall distribution of netizens' emotions, and comment hot words.

At present, our experiment has tapped 1,526 hot topics and 190,242 hot comments. However, there are still problems in our experiment. The experimental results show that the sentiment discrimination model of this system is more sensitive to the positive elements in the sentence. Although we have considered the influence of adverbials in the sentence on the pessimistic tone, the effect is limited. We have not given reliable experiments to prove that the system's decision process is equally effective in other corpus environments. And this model has not considered the emotional impact of the expression package on the comment text.

Our future work includes the following 3 aspects. First, we will consider the effect of expression packs on emotional expression. The use of emoticons in online language can greatly enhance the emotional color, but our data mining process artificially removes these factors. Netizens of different ages have different understandings of the same set of emoticons, and we cannot obtain comments through normal channels. The age information of the person. Second, we will continue to expand the Chinese sentiment dictionary of online public opinion related to the spread of the virus. Online language is changing with each passing day, and different online groups each have a set of online slang. Traditional dictionaries cannot recognize these words very accurately. A good dictionary can increase the efficiency of word segmentation and improve the generalization ability, accuracy, and recall rate of the model. Finally, we will optimize the crawler framework, moderately increase the request frequency, and improve the overall crawling efficiency. To ensure that the crawler still has enough information crawling ability when it touches some sensitive information, we even introduced a scheme to disguise anonymous proxy access, but it greatly slowed down the data crawling speed.

**Funding Statement:** This work was supported by the Hainan Provincial Natural Science Foundation of China [2019RC041, 2019RC098]; National Natural Science Foundation of China [61762033]; Hainan University Doctor Start Fund Project [kyqd1328]; Hainan University Youth Fund Project [qnjj1444]; Ministry of education humanities and social sciences research program fund project(19YJA710010); The Opening Project of Shanghai

Trusted Industrial Control Platform.

**Conflicts of Interest:** The authors declare that they have no conflicts of interest to report regarding the present study.